\documentclass[a4paper]{article}

\usepackage{INTERSPEECH2022}
\usepackage{pifont}
\usepackage{makecell}
\newcommand{\tabincell}[2]{\begin{tabular}
{@{}#1@{}}#2\end{tabular}}
\newcommand{\xmark}{\ding{55}}

\usepackage{multirow}
\usepackage{hhline}
\usepackage{graphicx,float}
\usepackage{subfigure} 
\usepackage{color}
\usepackage{cite}

\makeatletter
\renewcommand{\maketag@@@}[1]{\hbox{\m@th\normalsize\normalfont#1}}%
\makeatother

\title{Confidence Score Based Conformer Speaker Adaptation for Speech Recognition}
\name{Jiajun Deng$^{1*}$\thanks{$*$ Equal contribution}, Xurong Xie$^{2*}$, Tianzi Wang$^1$, Mingyu Cui$^1$, Boyang Xue$^1$, Zengrui Jin$^1$,\\ Mengzhe Geng$^1$, Guinan Li$^1$, Xunying Liu$^1$, Helen Meng$^1$}
\address{
  $^1$The Chinese University of Hong Kong, Hong Kong SAR, China\\
  $^2$Institute of Software, Chinese Academy of Sciences, China
  }
\email{\footnotesize{\{jjdeng,tzwang,mycui,byxue,zrjin,mzgeng,gnli,xyliu,hmmeng\}@se.cuhk.edu.hk, xurong@iscas.ac.cn}}

\begin{document}
\bstctlcite{IEEEexample:BSTcontrol}

\maketitle
\begin{abstract}
A key challenge for automatic speech recognition (ASR) systems is to model the speaker level variability. In this paper, compact speaker dependent learning hidden unit contributions (LHUC) are used to facilitate both speaker adaptive training (SAT) and test time unsupervised speaker adaptation for state-of-the-art Conformer based end-to-end ASR systems. The sensitivity during adaptation to supervision error rate is reduced using confidence score based selection of the more “trustworthy” subset of speaker specific data. A confidence estimation module is used to smooth the over-confident Conformer decoder output probabilities before serving as confidence scores. The increased data sparsity due to speaker level data selection is addressed using Bayesian estimation of LHUC parameters. Experiments on the 300-hour Switchboard corpus suggest that the proposed LHUC-SAT Conformer with confidence score based test time unsupervised adaptation outperformed the baseline speaker independent and i-vector adapted Conformer systems by up to 1.0\%, 1.0\%, and 1.2\% absolute (9.0\%, 7.9\%, and 8.9\% relative) word error rate (WER) reductions on the NIST Hub5’00, RT02, and RT03 evaluation sets respectively. Consistent performance improvements were retained after external Transformer and LSTM language models were used for rescoring.

\end{abstract}

\section{Introduction}
%\vspace{-0.1cm}
Recently, convolution-augmented Transformer (Conformer) \cite{gulati2020conformer,guo2021recent} end-to-end (E2E) automatic speech recognition (ASR) systems have been successfully applied to a wide range of task domains. Compared with the earlier forms of Transformer models \cite{Dong2018SpeechTransformerAN,Karita2019ACS}, Conformer benefits from a combined use of self-attention and convolution structures that are designed to learn both the longer range, global and local contexts.

\begin{figure*}[htbp]
    \centering
    \includegraphics[scale=0.38]{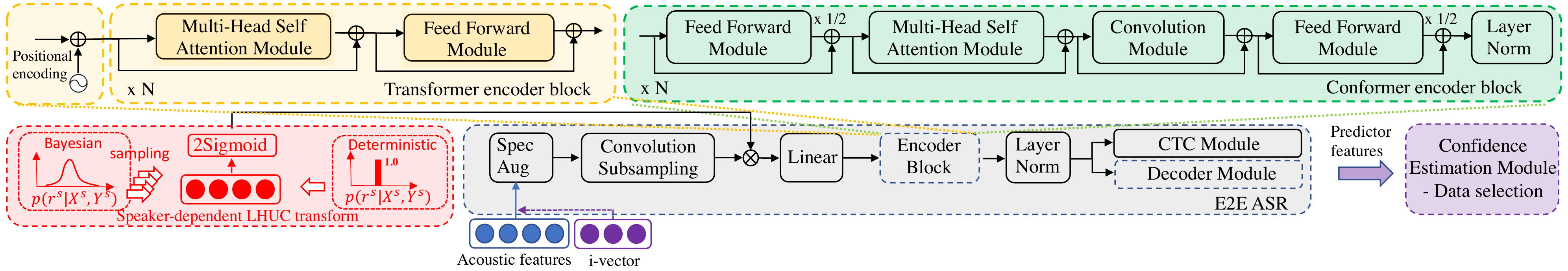}
    \setlength{\abovecaptionskip}{2pt plus 1pt minus 1pt}
    \caption{An example Conformer E2E architecture is shown in the grey box at the bottom centre. Its encoder internal components are also shown in the green coloured box on the top right. The standard Transformer encoder without the additional Convolutional module is shown in the light yellow coloured box, top left. Bayesian estimation or fixed value, deterministic LHUC speaker adaptation are shown on the left and right of the red coloured box in the bottom left corner.}
    \label{fig:figure1_arch}
    \vspace{-0.5cm}
\end{figure*}

A key challenge for ASR systems, including those based on Conformers, is to model the systematic and latent variation in speech. A major source of such variability is attributable to speaker level characteristics representing factors such as accent and idiosyncrasy, or physiological differences manifested in, for instance, age or gender. Speaker adaptation techniques for current ASR systems can be characterized into several categories: 1) Auxiliary speaker embedding-based approaches utilizing a compact vector (e.g., i-vector \cite{saon2013speaker}) to represent speaker-dependent (SD) characteristics \cite{saon2013speaker,senior2014improving,abdel2013fast,xue2014direct}. The SD embedding vectors are then fed into the system as auxiliary features and to augment standard acoustic front-ends. 2) Feature transformation-based methods, for example, feature-space maximum likelihood linear regression \cite{gales1998maximum}, and vocal tract length normalization\cite{lee1996speaker,uebel1999investigation}, aim to produce speaker-independent (SI) features by applying feature transform in the acoustic front-ends. 3) Model-based adaptation that uses separate SD model parameters to represent speaker variability, for example, learning hidden unit contributions (LHUC) \cite{Swietojanski2014LearningHU,Swietojanski2016LearningHU,Xie2021BayesianLF}, parameterized hidden activation functions \cite{Zhang2016DNNSA,Huang2017BayesianUB} and various forms of linear transforms that are applied to different neural network layers \cite{Neto1995SpeakeradaptationFH,Gemello2007LinearHT,Li2010ComparisonOD,Zhang2015ParameterisedSA,Zhao2016LowrankPD,Tan2016ClusterAT}. These prior researches on speaker adaptation were predominantly conducted in the context of traditional hybrid HMM-DNN based ASR systems. 

In contrast, very limited previous researches on speaker adaptation have been conducted for E2E ASR systems. Among these, regularization-based methods were exploited to improve the generalization performance of connectionist temporal classification (CTC) based \cite{Li2018SpeakerAF} and attention-based E2E models \cite{Meng2019SpeakerAF}. Auxiliary speaker embedding vectors based on i-vector \cite{Tuske2021OnTL}, or extracted from sequence summary network \cite{Delcroix2018AuxiliaryFB} and speaker-aware modules \cite{Fan2019SpeakerAwareS,Zhao2020SpeechTW} were incorporated into attention-based encoder-decoder or conventional, non-convolution augmented Transformer models. Speaker adaptation of Conformer RNN-T models was investigated in \cite{Huang2021RapidSA}. 
%% To the best of our knowledge, there has been \textcolor{red}{limited} model-based speaker adaptation techniques designed for Conformer E2E ASR systems to date. 

Efforts on developing model-based speaker adaptation techniques for Conformer models are confronted with several challenges. First, the often limited amounts of speaker specific data requires highly compact SD parameter representations to be used to mitigate the risk of overfitting during adaptation. Second, model adaptation techniques such as LHUC \cite{Swietojanski2014LearningHU,Swietojanski2016LearningHU} operate within a multi-pass decoding framework. An initial recognition pass produces the initial speech transcription to serve as the supervision for the subsequent SD parameter estimation and re-decoding of speech in a second pass. As a result, SD parameter estimation is sensitive to the underlying supervision error rate. This issue is particularly prominent with E2E systems that directly learn the surface word or token sequence labels in the adaptation supervision, as opposed to latent phonetic targets in hybrid ASR systems.

In order to address these issues, compact speaker dependent LHUC, for example, of 5120-dimensions, are used to facilitate both speaker adaptive training (SAT) \cite{Anastasakos1996ACM} and test time unsupervised speaker adaptation. The sensitivity to supervision quality is reduced using confidence score based selection of the more “trustworthy” subset of speaker specific data. In Conformer models, the attention-based encoder-decoder model architecture and auto-regressive decoder component design both of which utilize the full history context, make the use of conventional lattice-based recognition hypotheses representation for word posterior and confidence score estimation non-trivial \cite{evermann2000posterior,Mangu2000FindingCI}. To this end, a lightweight token-level confidence estimation module (CEM) is incorporated to smooth the over-confident \cite{Hendrycks2017ABF,Li2021ConfidenceEF,Liu2021UtteranceLevelNC} Conformer decoder output probabilities and improve their reliability as confidence scores. The data sparsity issue further increased by the use of confidence score-based speaker level data selection is addressed using Bayesian estimation \cite{Xie2019BLHUCBL,Xie2021BayesianLF} of LHUC parameters.

Experiments on the 300-hour Switchboard corpus suggest that the proposed confidence score-based selection of speaker level subset data of varying quantities produced adaptation performance comparable to word error rate (WER) based selection. The proposed LHUC-SAT Conformer with confidence score-based test time unsupervised adaptation outperformed the baseline SI and i-vector adapted Conformer systems by up to 1.0\%, 1.0\%, and 1.2\% absolute (9.0\%, 7.9\%, and 8.9\% relative) WER reductions on the NIST Hub5’00, RT02, and RT03 evaluation sets respectively. Consistent performance improvements were retained after external Transformer and LSTM language models were used for rescoring. 

The rest of this paper is organized as follows. Section 2 reviews the conventional Transformer and Conformer ASR systems. LHUC-SAT Conformer training and confidence score-based Bayesian LHUC test time unsupervised adaptation are proposed in Section3. Section 4 presents the experiments and results. Finally, the conclusions are drawn in Section 5. 

\vspace{-0.2cm}
\section{Conformer ASR Architecture}
\vspace{-0.1cm}

The convolution-augmented Transformer (Conformer) model follows the architecture proposed in \cite{gulati2020conformer,guo2021recent}. It consists of a Conformer encoder and a Transformer decoder. The Conformer encoder is based on a multi-blocked stacked architecture. Each encoder block includes the following components in turn: a position wise feed-forward (FFN) module, a multi-head self-attention (MHSA) module, a convolution (CONV) module and a final FFN module at the end. Among these, the CONV module further consists of in turn: a 1-D pointwise convolution layer, a gated linear units (GLU) activation \cite{Dauphin2017LanguageMW}, a second 1-D pointwise convolution layer followed a 1-D depth-wise convolution layer, a Swish activation and a final 1-D pointwise convolution layer. Layer normalization (LN) and residual connections are also applied to all encoder blocks. An example Conformer architecture is shown in Figure \ref{fig:figure1_arch}.

For both the Transformer and Conformer models, the following multitask criterion interpolation between the CTC and attention error costs \cite{Watanabe2017HybridCA} is used in training.
\begin{equation}
%\setlength{\abovedisplayskip}{0pt plus 1pt minus 1pt}
%\vspace{-0.1cm}
{\cal{L}}=(1-\lambda) {{\cal{L}}_{att}} + \lambda{{\cal{L}}_{ctc}},
\label{enq:loss_combine}
\end{equation}
where $\lambda$ is empirically set as 0.2 during training and fixed throughout the experiments of this paper.

%\vspace{-0.2cm}
\vspace{-0.1cm}
\section{Conformer Speaker Adaptation}
\vspace{-0.1cm}
Confidence score based LHUC Conformer adaptation and Bayesian estimation are presented in this section.
%\vspace{-0.2cm}
\subsection{LHUC}
The key idea of LHUC \cite{Swietojanski2014LearningHU,Swietojanski2016LearningHU,Xie2021BayesianLF} and the related parameterised adaptive activation functions \cite{Zhang2016DNNSA,Huang2017BayesianUB} is to modify the amplitudes of hidden unit activations of hybrid DNN, or Conformer models as considered in this paper, for each speaker using a SD transformation. This can be parameterized by using activation output scaling vectors. Let ${\bm{r}}^{l,s}$ denote the SD parameters for speaker $s$ in the $l$-th hidden layer, the hidden layer output can be given as 
%\vspace{-0.1cm}
\begin{equation}
%\vspace{-0.2cm}
\setlength{\abovedisplayskip}{0pt plus 1pt minus 1pt}
{\bm{h}}^{l,s}=\xi(\bm{r}^{l,s})\odot {\bm{h}}^{l},
\setlength{\belowdisplayskip}{0pt plus 1pt minus 1pt}
% \vspace{-0.2cm}
\end{equation}
where $\odot$ denotes the Hadamard product operation, ${\bm{h}}^{l}$ is the hidden vector after a non-linear activation function in the $l$-th layer, and $\xi(\bm{r}^{l,s})$ is the scaling vector parametrized by ${\bm{r}}^{l,s}$. In this paper, $\xi(\cdot)$ is the element-wise 2Sigmoid$(\cdot)$ function with range $(0, 2)$. Given the adaptation data ${\cal{D}}^{s}=\{{\bm{X}}^{s},{\bm{Y}}^{s}\}$ for speaker $s$, ${\bm{X}}^{s}$ and ${\bm{Y}}^{s}$ stand for the acoustic features and the corresponding supervision token sequences, respectively. The estimation of SD parameters ${\bm{r}}^s$ can be obtained by minimizing the loss in Eq. (\ref{enq:loss_combine}), which is given by
%\vspace{-0.1cm}
\begin{equation}%\hspace{-1.8mm}
\hat{{\bm{r}}}^s\!=\!\arg\min\limits_{{\bm{r}}^s}\{\lambda_1\!\log p_{a}\!({\bm Y}^{s}\!|\!{\bm X}^s,\!{\bm r}^s) \!-\!\lambda\!\log p_{c}({\bm Y}^{s}\!|\!{\bm X}^s\!,\!{\bm r}^s)\}
\vspace{-0.2cm}
\end{equation}
where $\lambda_1=\lambda-1$, $p_{a}({\bm Y}^{s}|{\bm X}^s, {\bm r}^s)$ and $p_{c}({\bm Y}^{s}|{\bm X}^s, {\bm r}^s)$ are the attention-based and CTC-based likelihood probabilities, respectively. The supervisions ${\bm{Y}}^{s}$ that are not available during unsupervised test time adaptation can be obtained by initially decoding the corresponding utterances using a baseline SI model, before serving as the target token labels in the subsequent adaptation and re-decoding stage. As a result, SD parameter estimation is sensitive to the underlying supervision error rate.
\vspace{-0.2cm}
\subsection{Confidence Score Based LHUC Adaptation}
% \vspace{-0.2cm}
To mitigate the risk of adaptation using speaker level data with incorrect token labels, one possible solution considered here is to exploit confidence score-based selection of a more “trustworthy”, accurate and less erroneous subset of speaker specific data. The use of confidence scores featured widely in both speaker adaptation \cite{Anastasakos1998TheUO,Wallhoff2000FramediscriminativeAC,Liu2013LanguageMC} and unsupervised training \cite{Zavaliagkos1998UsingUT,Kemp1999UnsupervisedTO,Yu2010UnsupervisedTA} of traditional HMM-based ASR systems.

The efficacy of confidence score-based data selection crucially depends on its correlation with the WER. Conventional HMM-based ASR systems use lattice-based recognition hypotheses representation for word or token posterior and confidence score estimation \cite{evermann2000posterior,Mangu2000FindingCI,Seigel2011CombiningIS}. However, the attention-based encoder-decoder model architecture and auto-regressive decoder component design utilizing the full history context used in Conformer, and other related E2E models \cite{Prabhavalkar2021LessIM}, lead to their difficulty in deploying effective beam search in lattice-based decoding. An alternative solution is to use the Conformer decoder output probabilities as confidence scores. However, these have been found to be over-confident in previous research \cite{Li2021ConfidenceEF,Liu2021UtteranceLevelNC}. To this end, a lightweight token-level CEM is used in this paper to further smooth the over-confident Conformer decoder output probabilities and improve their reliability as confidence scores. 

The form of CEM considered in this paper is a lightweight binary classification model using a 3-layer residual FFN that can be easily configured on top of a well-trained E2E ASR system such as Conformer. For each output token, its corresponding hidden state extracted from the last decoder layer and linear outputs corresponding to top-10 Softmax probabilities are concatenated as the input features fed into the CEM, which then produces a smoothed confidence score for the token. An utterance-level confidence score can then be obtained by averaging the confidence scores of all tokens within an utterance. The alignments between the hypothesis supervision and the ground truth reference can be obtained from the edit distance computation, which can then be used as the target labels for CEM training, where the correct tokens are assigned with the value of one while the substituted or inserted tokens are assigned with zero. 
 
\vspace{-0.2cm}
\subsection{Bayesian LHUC Adaptation}
\vspace{-0.1cm}
Data selection further reduces the amount of limited speaker level adaptation data, and increases the sparsity issue. This leads to uncertainty in SD parameters during the standard LHUC adaptation performing deterministic parameter estimation. The solution adopted in this paper uses Bayesian learning \cite{Xie2019BLHUCBL} by modelling SD parameters’ uncertainty, and the following predictive distribution is utilized in LHUC adaptation. The prediction over the test utterance ${\tilde{\bm X}}^s$ is given by
\vspace{-0.3cm}
\begin{equation}
p(\tilde{\bm Y}^s|\tilde{\bm X}^s,{\cal D}^s)=\int{p(\tilde{\bm{Y}}^{s}|\tilde{\bm{X}}^{s},{\bm r}^s)p({\bm r}^s|{\cal D}^s)d{\bm r}^s},
\label{eq:posterior}
\vspace{-0.3cm}
\end{equation}
where ${\tilde{\bm Y}}^s$ is the predicted token sequence and $p({\bm r}^s|{\cal D}^s)$ is the SD parameters posterior distribution. The true posterior distribution can be approximated with a variational distribution $q({\bm r}^s)$ by minimising the following bound of hybrid attention plus CTC loss marginalisation over the adaptation data set ${\cal D}^s$,
\begin{align}
\vspace{-1cm}
&\lambda_1\log\int{p_{a}({\bm Y}^{s}, {\bm r}^s|{\bm X}^s)d{\bm r}^s}
-\lambda\log\int{p_{c}({\bm Y}^{s}, {\bm r}^s|{\bm X}^s)d{\bm r}^s} \nonumber \\
\leq& \int{q({\bm r}^s)(\lambda_1\log p_{a}({\bm Y}^{s}|{\bm X}^s\!,\!{\bm r}^s) -\lambda\log p_{c}({\bm Y}^{s}|{\bm X}^s\!,\!{\bm r}^s))d{\bm r}^s} \nonumber \\
&+KL(q({\bm r}^s)||p({\bm r}^s))\triangleq {\cal L}_1 + {\cal L}_2,
% \\ 
% +&\lambda \log\int{\exp\{{\cal L}_{ctc}({\cal D}^s; {\bm r}^s)}\}p({\bm r}^s)d{\bm r}^s  
\vspace{-0.4cm}
\end{align}
where $p({\bm r}^s)$ is the prior distribution of SD parameters, and KL$(\cdot)$ is the KL divergence. For simplicity, both $q({\bm r}^s)={\cal N}({\bm \mu}, {\bm \sigma}^2)$ and $p({\bm r}^s)={\cal N}({\bm \mu}_r, {\bm \sigma}_r^2)$ are assumed to be normal distributions. The first term ${\cal L}_1$ is approximated with Monte Carlo sampling method, which is given by
\begin{equation}
%\hspace{-0.5mm}
\vspace{-0.2cm}
{\cal L}_1\!\approx\! \frac{1}{N}\!\sum\limits_{k=1}^{N}\!{\lambda_1\!\log p_{a}\!({\bm Y}^{s}|{\bm X}^s\!,\!{\bm r}_k^s)\!\!-\!\!\lambda\log p_{c}({\bm Y}^{s}|{\bm X}^s\!,\!{\bm r}_k^s)}
\label{eq:loss1}
%\vspace{-0.1cm}
\end{equation} % 
where ${\bm r}_k^s={\bm \mu}+{\bm \sigma}\odot {\bm \epsilon}_k$ and ${\bm \epsilon}_k$ is the $k$-th Monte Carlo sampling value drawn from the standard normal distribution. The KL divergence ${\cal L}_2$ can be explicitly calculated as
\begin{equation}
\vspace{-0.2cm}
\mathcal{L}_2=\frac{1}{2}\sum_{i}{(\frac{\sigma_i^2+(\mu_i-\mu_{r,i})^2}{\sigma_{r,i}^2}+2\log\frac{\sigma_{r,i}}{\sigma_{i}}-1}),
%\vspace{-0.2cm}
\end{equation}
where $\{\mu_{r,i},\sigma_{r,i}\}$ and $\{\mu_i,\sigma_i\}$ are the $i$-th elements of vectors $\{{\bm \mu}_r,{\bm \sigma}_r\}$ and $\{{\bm \mu},{\bm \sigma}\}$, respectively.

\textbf{Implementation details} over several crucial settings are: 1) Based on empirical evaluation, the LHUC layer is applied to the hidden output of the convolution subsampling module, as shown in Figure \ref{fig:figure1_arch} (bottom, left). Alternative locations\footnote{LHUC scaling is applied to the hidden outputs of the convolutional, linear, or attention layer, and various combinations of these locations.} to apply the LHUC transforms in practice leads to performance degradation in comparison. 2) The LHUC prior distribution is modelled by a standard normal distribution ${\cal N}(\bf 0, 1)$. 3) In Bayesian learning, only one parameter sample is drawn in Eq. (\ref{eq:loss1}) during adaptation to ensure that the computational cost is comparable to that of standard LHUC adaptation. 4) The predictive inference integral in Eq. (\ref{eq:posterior}) is efficiently approximated by the expectation of the posterior distribution as  $p(\tilde{\bm{Y}}^{s}|\tilde{\bm{X}}^{s},{{\mathbb{E}}[{\bm r}^s|{\cal D}^s]})$.

\vspace{-0.2cm}
\section{Experiments}
%\vspace{-0.1cm}
In this section, the proposed confidence score-based LHUC speaker adaptation method is investigated for unsupervised test adaptation of both SI and speaker adaptively trained SAT (interleaving update of TDNN and SD LHUC parameters, following \cite{Swietojanski2016LearningHU}) ASR systems on the 300-hr Switchboard task. 
\vspace{-0.1cm}
\subsection{Experimental Setup}
\%vspace{-0.1cm}
The \textbf{Switchboard-1} corpus \cite{Godfrey1992SWITCHBOARDTS} with 286-hr speech collected from 4804 speakers is used for training. The NIST \textbf{Hub5’00}, \textbf{RT02} and \textbf{RT03} evaluation sets containing 80, 120 and 144 speakers, and 3.8, 6.4 and 6.2 hours of speech respectively were used. 80-dim Mel-filter bank plus 3-dim pitch parameters were used as input features. The ESPnet recipe \cite{guo2021recent} configured Conformer model contains 12 encoder and 6 decoder layers, where each layer is configured with 4-head attention of 256-dim, and 2048 feed forward hidden nodes. Byte-pair-encoding (BPE) tokens of size 2000 were served as the decoder outputs. The convolution subsampling module contains two 2-D convolutional layers with stride 2. SpecAugment \cite{Park2019SpecAugmentAS} was applied in both SI and SAT Conformer training. The Noam optimizer’s initial learning rate was 5.0. The dropout rate was set as 0.1, and model averaging was performed over the last ten epochs. The confidence score estimation module is a 3-layer residual feedforward DNN with a hidden dimension of 64 with batch normalization and dropout applied in training. Decoding outputs were further rescored using log-linearly interpolated external Transformer and Bidirectional LSTM language models (LMs) trained on the Switchboard and Fisher transcripts using cross-utterance contexts \cite{Sun2021TransformerLM}. 

\vspace{-0.2cm}
\begin{table}[H]
\centering
\setlength{\abovecaptionskip}{1pt plus 1pt minus 1pt}
\setlength{\belowcaptionskip}{0pt plus 1pt minus 1pt}
\caption{Performance (WER\%) of SI and LHUC-SAT trained Transformer and Conformer systems before and after test time unsupervised LHUC adaptation evaluated on Hub5’00, RT02 and RT03. “CHE”, “SWBD”, “FSH” and “O.V.” stand for CallHome, Switchboard, Fisher and Overall respectively. $\dagger$ denotes a statistically significant (MAPSSWE, $\alpha$=0.05) difference obtained over baseline SI systems (sys. 1, 4).}
% \vspace{10pt}
\label{tab:table1}
\resizebox{\columnwidth}{!}{
\begin{tabular}{c|l|ccc|cccc|ccc} %
	\hline\hline
    \multirow{2}{*}{Sys.} & \multirow{2}{*}{Model}  & \multicolumn{3}{c|}{Hub5'00 } & \multicolumn{4}{c|}{RT02} & \multicolumn{3}{c}{RT03} 
    \\   \cline{3-12}
	& & CHE & SWBD & O.V. & SWBD1 & SWBD2 & SWBD3 & O.V. & FSH & SWBD & O.V.  \\ \hline\hline
	1 & Transformer & 17.3 & 9.1 & 13.2 & 10.5 & 15.1 & 18.2 & 14.9 & 12.1 & 19.0 & 15.7 \\ 
	2 & + LHUC & 16.8 & 9.1 & 13.0 & 10.5 & 15.0 & 17.9 & 14.7 & 12.0 & 18.9 & 15.6 \\ 
	3 & + LHUC-SAT  & ${16.1}^{\dagger}$ & ${8.5}^{\dagger}$ & ${12.3}^{\dagger}$ & 10.1 & ${14.4}^{\dagger}$ & ${17.4}^{\dagger}$ & ${14.2}^{\dagger}$ & ${11.4}^{\dagger}$  & ${17.7}^{\dagger}$ & ${14.6}^{\dagger}$ \\ \hline\hline
	4 & Conformer & 15.0 & 7.3 & 11.1 & 8.7 & 12.9 & 15.6 & 12.6 & 10.4 & 16.4 & 13.5 \\ 
	5 & + LHUC-SAT & ${\textbf {{13.8}}^{\dagger}}$ & ${\textbf{7.1}}$ & ${\textbf{{10.5}}^{\dagger}}$ & ${\textbf{8.6}}$ & ${\textbf{{12.3}}^{\dagger}}$ & ${\textbf{{14.7}}^{\dagger}}$ & ${\textbf{{12.0}}^{\dagger}}$ & ${\textbf{{10.0}}^{\dagger}}$  & ${\textbf{{15.7}}^{\dagger}}$ & ${\textbf{{12.9}}^{\dagger}}$ \\ %\cline{3-12}
	\hline \hline

\end{tabular} 
}
\end{table}%}
\begin{figure*}[htbp]
\centering
%\subfigure[pic1.]{
\begin{minipage}[t]{0.3\linewidth}
\centering
\includegraphics[width=2.0in]{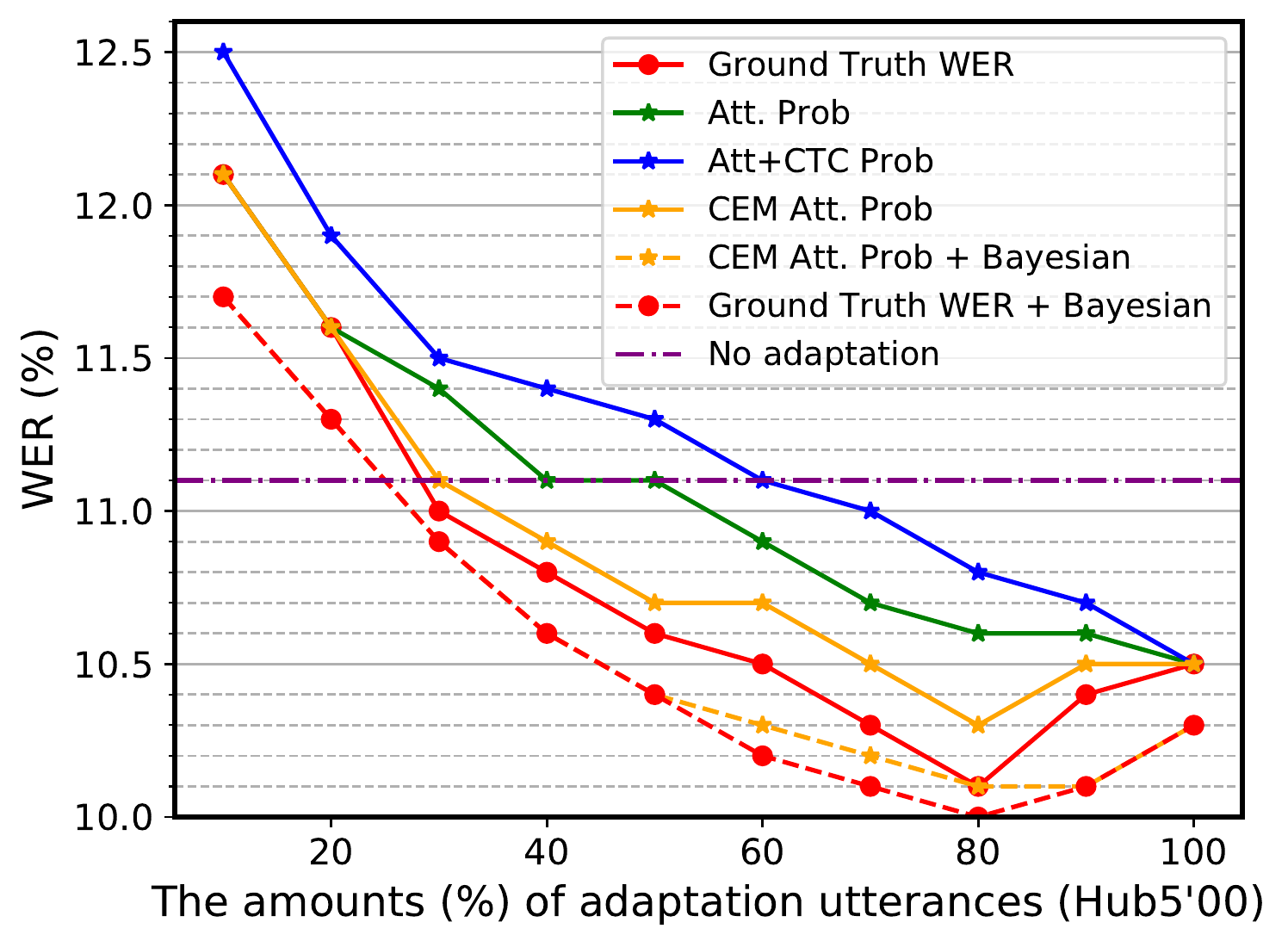}
%\caption{fig1}
\end{minipage}%
%}%
%\subfigure[]{
\begin{minipage}[t]{0.3\linewidth}
\centering
\includegraphics[width=2.0in]{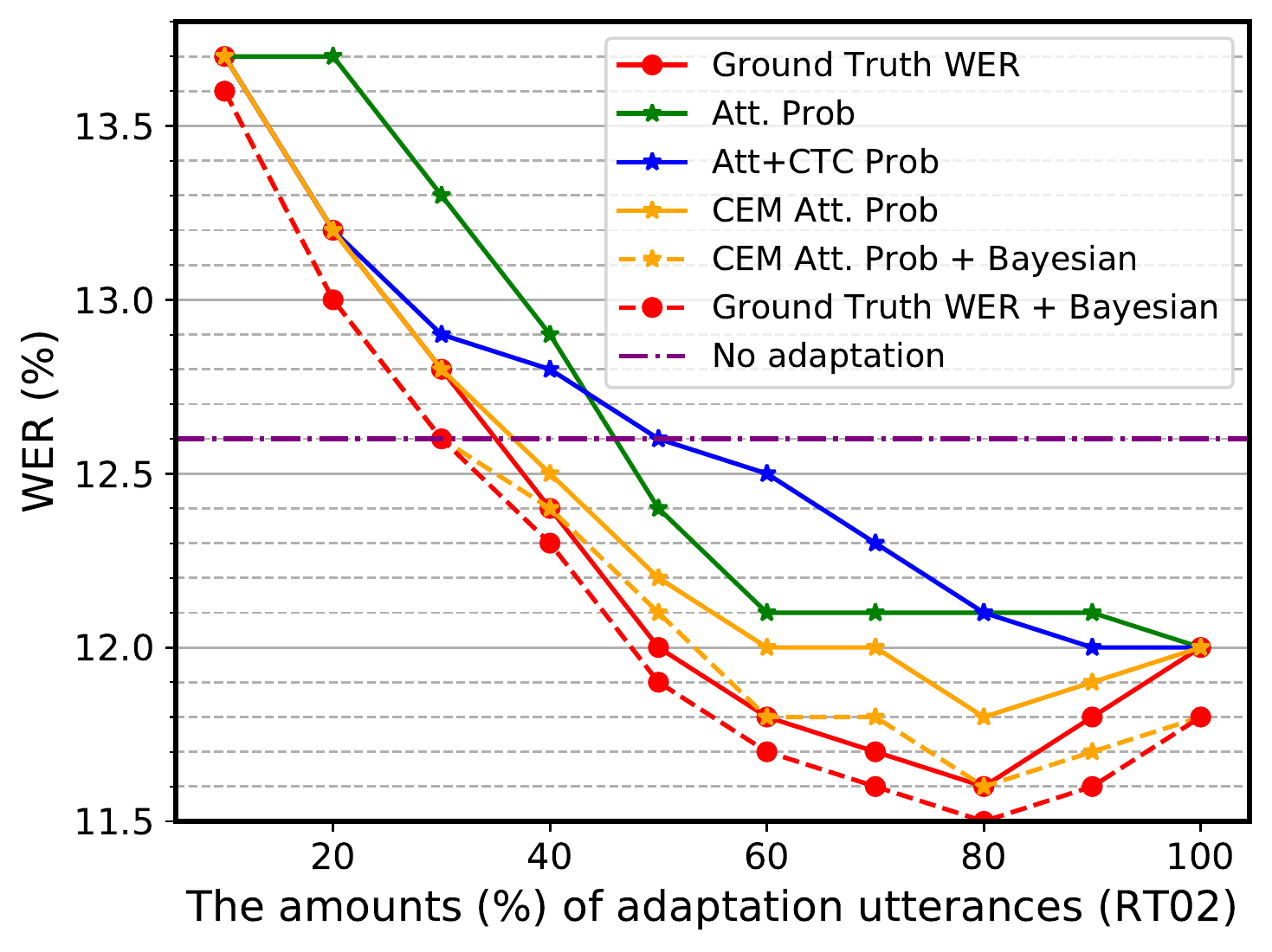}
%\caption{fig2}
\end{minipage}%
%}%
%\subfigure[pic3.]{
\begin{minipage}[t]{0.28\linewidth}
\centering
\includegraphics[width=2.0in]{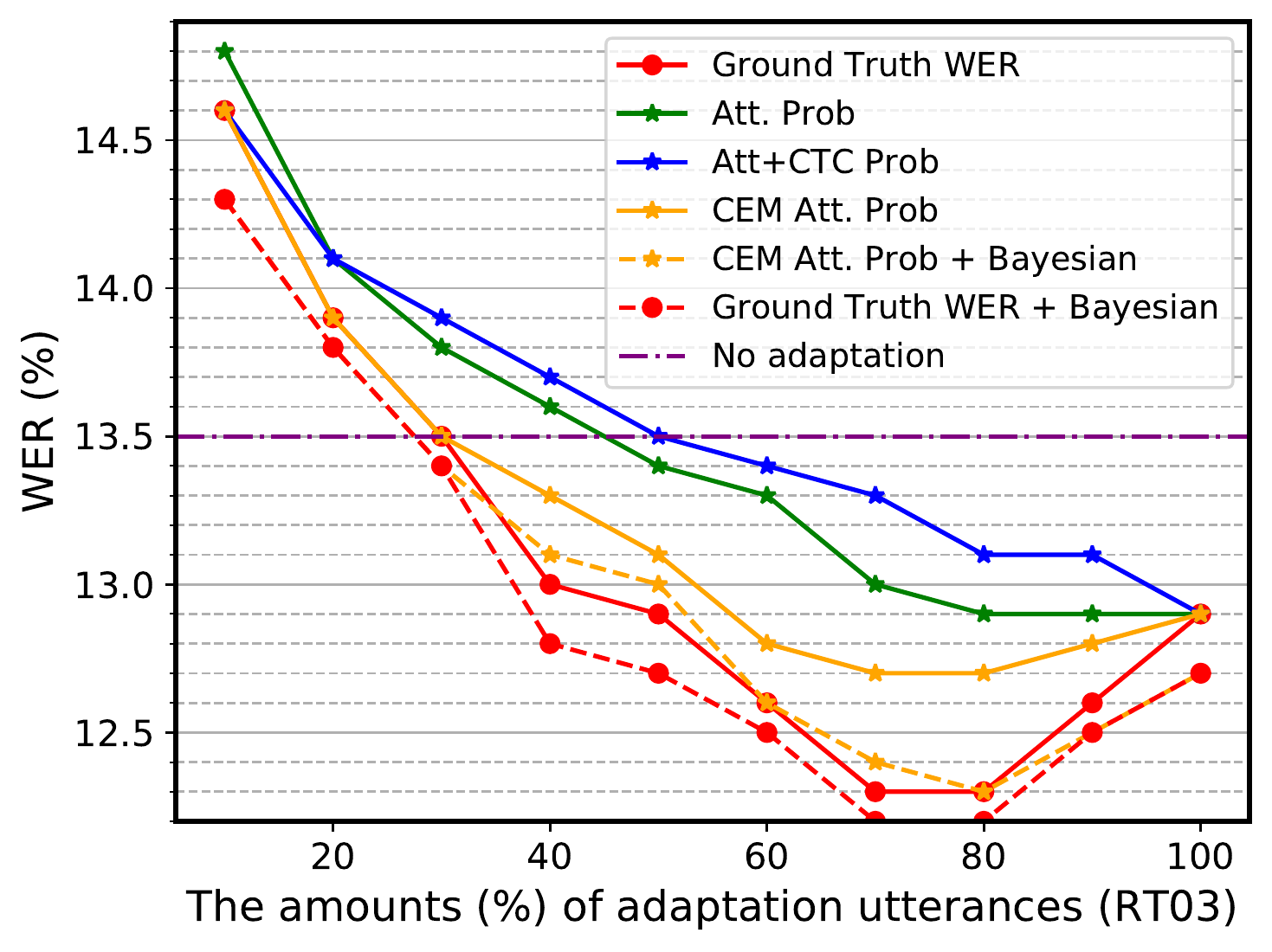}
%\caption{fig2}
\end{minipage}
%}%
\centering
\setlength{\abovecaptionskip}{1pt plus 1pt minus 1pt}
\caption{Performance (WER\%) of adapted LHUC-SAT Conformer systems using various amounts of adaptation data selected using different methods (top down over legends): ground truth WER “Ground Truth WER”; original Conformer decoder attention Softmax output probabilities “Att. Prob”, or their combination with CTC “Att+CTC Prob”; the CEM smoothed attention scores “CEM Att. Prob”. “+Bayesian” denotes Bayesian LHUC test adaptation.}
\vspace{-0.2cm}

\label{fig:fig1}
\end{figure*}
%\vspace{-0.5cm}
\begin{table*}[htbp]
\centering
\setlength{\abovecaptionskip}{1pt plus 1pt minus 1pt}
\caption{Speaker adapted performance (WER\%) of LHUC-SAT Conformer systems with/without confidence score-based data selection and Bayesian LHUC adaptation on the Hub5’00, RT02 and RT03 data sets, before and after external Transformer plus LSTM LM rescoring. $\dagger$ denotes a statistically significant (MAPSSWE, $\alpha$=0.05) WER difference over the baseline SI systems (sys. 1, 6).}
\label{tab:table2}
\resizebox{2.0\columnwidth}{!}{
\begin{tabular}{c|c|c|c|c|ccc|cccc|ccc} %
	\hline\hline
    \multirow{2}{*}{Sys.} & 
    \multirow{2}{*}{\tabincell{c}{Speaker\\Adaptation}}  &
    \multirow{2}{*}{\tabincell{c}{Adaptation \\Parameters}} &
    \multirow{2}{*}{\tabincell{c}{Data\\Selection}} & 
    \multirow{2}{*}{\tabincell{c}{Language\\Model}} &   \multicolumn{3}{c|}{Hub5'00 } & \multicolumn{4}{c|}{RT02} & 
    \multicolumn{3}{c}{RT03} 
    \\   \cline{6-15}
	& & &  & & CHE & SWBD & O.V. & SWBD1 & SWBD2 & SWBD3 & O.V. & FSH & SWBD & O.V.  \\ \hline\hline
	1 & \xmark & \xmark & \xmark & \multirow{5}{*}{\xmark}& 15.0 & 7.3 & 11.1 & 8.7 & 12.9 & 15.6 & 12.6 & 10.4 & 16.4 & 13.5\\ 
	2 &  i-vector\footnotemark[2] & \xmark & \xmark &  & 15.8 & 7.8 & 11.8 & 9.3 & 13.6 & 16.3 & 13.3 & 10.9 & 17.3 & 14.2\\ 

	3 &  LHUC-SAT & Deterministic & \xmark & & ${13.8}^{\dagger}$ & 7.1 & ${10.5}^{\dagger}$ & 8.6 & ${12.3}^{\dagger}$ & ${14.7}^{\dagger}$ & ${12.0}^{\dagger}$ & ${10.0}^{\dagger}$  & ${15.7}^{\dagger}$ &${12.9}^{\dagger}$ \\  
	4 &  LHUC-SAT & Deterministic & \checkmark & & ${13.7}^{\dagger }$ & ${6.9}^{\dagger }$ & ${10.3}^{\dagger }$ & ${\textbf {8.3}^{\dagger}}$ & ${12.0}^{\dagger }$ & ${14.5}^{\dagger }$ & ${11.8}^{\dagger }$ & ${9.8}^{\dagger }$ & ${15.4}^{\dagger }$ & ${12.7}^{\dagger }$\\ 
	5 &  LHUC-SAT & Bayesian & \checkmark & & ${\textbf {13.4}^{\dagger }}$ & ${\textbf {6.8}^{\dagger }}$& ${\textbf {10.1}^{\dagger}}$ & ${\textbf {8.3}^{\dagger}}$ & ${\textbf {11.9}^{\dagger}}$ &${\textbf {14.1}^{\dagger}}$ & ${\textbf {11.6}^{\dagger}}$ & ${\textbf {9.5}^{\dagger}}$ & ${\textbf {14.8}^{\dagger}}$ & ${\textbf {12.3}^{\dagger}}$\\ \hline \hline

	6 &  \xmark & \xmark & \xmark & \multirow{2}{*}{\tabincell{c}{Transformer + BiLSTM \\ (cross-utterance)}}& 14.2 & 6.8  & 10.5 & 8.5 & 12.3 & 14.8 & 12.1 & 9.6 & 15.3 & 12.6 \\ 
	7 &  LHUC-SAT & Bayesian & \checkmark & & ${\textbf {13.0}^{\dagger }}$& ${\textbf {6.3}^{\dagger }}$ & ${\textbf {9.6}^{\dagger }}$ & ${\textbf{8.1}^{\dagger }}$& ${\textbf {11.3}^{\dagger }}$ & ${\textbf {13.4}^{\dagger }}$ & ${\textbf {11.1}^{\dagger }}$ & ${\textbf {8.7}^{\dagger }}$ & ${\textbf {13.8}^{\dagger }}$ & ${\textbf {11.3}^{\dagger}}$\\ \hline \hline
\end{tabular} 
}
\vspace{-0.6cm}
\end{table*}%}
\footnotetext[2]{The Kaldi recipe configured 100-dim utterance-level i-vector features were concatenated to the acoustic features at the input layer.}
\vspace{-0.6cm}
\subsection{Performance of LHUC Adaptation}
%\vspace{-0.2cm}
The performance of non-Bayesian, deterministic parameter-based LHUC adaptation for both SI and LHUC-SAT trained Transformer and Conformer systems are shown in Table \ref{tab:table1} (sys. 2, 3, 5). For both Transformer and Conformer systems, LHUC-SAT with speaker adaptation consistently outperformed their respective SI baselines (sys. 3, 5 vs. sys. 1, 4).  In particular, the LHUC-SAT Transformer produced a statistically significant WER reduction of \textbf{1.1\%} absolute on the RT03 data over the SI baseline (sys. 3 vs. sys. 1). The LHUC-SAT Conformer system (sys. 5) produced the lowest WER and was used in the following experiments on confidence score-based adaptation. 
\vspace{-0.1cm}
\subsection{Performance of Confidence Based Adaptation}
%\vspace{-0.2cm}
The efficacy of the proposed confidence score-based speaker level data selection is first assessed via an ablation study on the three test sets, as shown in Figure \ref{fig:fig1}, where the WERs of adapted LHUC-SAT Conformer systems using various amounts of adaptation data selected using different methods are shown. The horizontal coordinates represent speaker level selected subsets of utterances ranked by their confidence scores (e.g., top 80 percentile) of varying sizes used for adaptation.

A general trend can be found that across all three test sets over various percentiles based subset selection, the proposed CEM smoothed Conformer attention Softmax output probabilities (orange lines) consistently produced better LHUC test adaptation performance than the comparable baseline selection using un-mapped, original Softmax output probabilities (green line). After Bayesian estimation is applied to mitigate the risk of over-fitting to subsets of speaker data, CEM smoothed confidence score-based Bayesian LHUC adaptation performance (orange dotted line) is comparable to that using the ground truth utterance level WER-based data selection (red dotted line). The best operating percentile point for smoothed confidence score-based data selection is approximately at 80\%. 

Using the resulting top 80 percentile selected speaker level data, the corresponding adapted performance of the LHUC-SAT Conformer systems with or without Bayesian LHUC estimation are further contrasted against the baseline SI and i-vector adapted Conformers in Table 2 (sys. 4, 5 vs. sys. 1, 2). Using both CEM smoothed confidence score in data selection and Bayesian learning in LHUC test adaptation, overall statistically significant WER reductions of \textbf{1.0\%}, \textbf{1.0\%} and \textbf{1.2\%} absolute (\textbf{9.0\%}, \textbf{7.9\%} and \textbf{8.9\%} relative) were obtained on the three test sets over the baseline SI Conformer (sys. 5 vs. sys. 1).

Similar performance improvements of \textbf{0.9\%}-\textbf{1.3\%} absolute WER reductions were retained after external Transformer plus LSTM LM rescoring (sys. 7 vs. sys. 6). Finally, the performance of the best confidence LHUC adapted LHUC-SAT Conformer system (sys. 7, Table \ref{tab:table2}) is further contrasted in Table \ref{tab:table3} with those of the state-of-the-art performance obtained on the same task using the most recent hybrid and end-to-end systems reported in the literature to demonstrate its competitiveness. 

\begin{table}[htbp]
\vspace{-0.2cm}
\centering
\setlength{\abovecaptionskip}{1pt plus 1pt minus 1pt}
\caption{Performance (WER\%) contrast between the best confidence score-based Bayesian LHUC-SAT conformer system and other state-of-the-art systems on the Hub5’00 and RT03 data.}
\label{tab:table3}
\resizebox{\columnwidth}{!}{
\begin{tabular}{c|l|l|ccc|ccc} %
	\hline\hline
    \multirow{2}{*}{ID} & 
    \multirow{2}{*}{System}  &
    \multirow{2}{*}{\# Param.} &
    \multicolumn{3}{c|}{Hub5'00 } & 
    \multicolumn{3}{c}{RT03} 
    \\   \cline{4-9}
	& &  & CHE & SWBD & O.V. & FSH & SWBD & O.V.  \\ \hline\hline
	1 & RWTH-2019 Hybrid\cite{Kitza2019CumulativeAF} & - & 13.5 & 6.7 & 10.2 & - & - & - \\ \hline
	2 & CUHK-2021 BLHUC-Hybrid\cite{Xie2021BayesianLF} & 15.2M &  12.7 & 6.7 & 9.7 & 7.9 & 13.4 & 10.7 \\ \hline
	3 & Google-2019 LAS\cite{Park2019SpecAugmentAS} & -  & 14.1 & 6.8 & (10.5) \\ \hline
	4 & \multirow{3}{*}{IBM-2020 AED \cite{tuske2020single}}  & 29M & 14.6 & 7.4 & (11.0) & - & - & - \\
	5 &   & 75M  & 13.4 & 6.8 & (10.1) & - & - & - \\
	6 &   & 280M & 12.5 & 6.4 & 9.5 & 8.4 & 14.8 & (11.7) \\ \hline
	7 & IBM-2021 CFM-AED\cite{Tuske2021OnTL} & 68M  & 11.2 & 5.5 & (8.4) & 7.0 & 12.6 & (9.9) \\ \hline
	8 & Salesforce-2020 \cite{wang2020investigation} & -  & 13.3 & 6.3 & (9.8) & - &- & 11.4 \\ \hline
	9 & \multirow{2}{*}{\tabincell{l}{\textbf{Baseline \{Table 2-Sys. 6\} (ours)}\\ \textbf{+Adapt. \{Table 2-Sys. 7\} (ours)}}} & 45M & 14.2 & 6.8  & 10.5 & 9.6 & 15.3 & 12.6 \\
	10 & & + 0.74M & 13.0 & 6.3 & 9.6 & 8.7 & 13.8 & 11.3  \\\hline \hline
\end{tabular} 
}
\vspace{-0.2cm}
\end{table}%}
\vspace{-0.3cm}
\section{Conclusions}
%\vspace{-0.1cm}
The paper proposed a confidence score-based unsupervised speaker adaptation approach for Conformer based speech recognition systems. The adaptation performance sensitivity to supervision quality is addressed using a smoothed confidence-based data selection scheme while the resulting increased data sparsity is mitigated using Bayesian learning. Experiments on the 300-hour Switchboard corpus demonstrated the efficacy of the proposed methods and produced up to 9.0\% relative word error rate reduction over the baseline Conformer system. Further research will focus on improving data selection and rapid on-the-fly adaptation approaches. 
%\vspace{-0.3cm}
\section{Acknowledgements}
%\vspace{-0.2cm}
This research is supported by Hong Kong RGC GRF grant No. 14200021, 14200218, 14200220, Innovation \& Technology Fund grant No. ITS/254/19 and ITS/218/21, National Key R\&D Program of China (2020YFC2004100).
\bibliographystyle{IEEEtran}

\bibliography{mybib}

% \begin{thebibliography}{9}
% \bibitem[1]{Davis80-COP}
%   S.\ B.\ Davis and P.\ Mermelstein,
%   ``Comparison of parametric representation for monosyllabic word recognition in continuously spoken sentences,''
%   \textit{IEEE Transactions on Acoustics, Speech and Signal Processing}, vol.~28, no.~4, pp.~357--366, 1980.
% \bibitem[2]{Rabiner89-ATO}
%   L.\ R.\ Rabiner,
%   ``A tutorial on hidden Markov models and selected applications in speech recognition,''
%   \textit{Proceedings of the IEEE}, vol.~77, no.~2, pp.~257-286, 1989.
% \bibitem[3]{Hastie09-TEO}
%   T.\ Hastie, R.\ Tibshirani, and J.\ Friedman,
%   \textit{The Elements of Statistical Learning -- Data Mining, Inference, and Prediction}.
%   New York: Springer, 2009.
% \bibitem[4]{YourName17-XXX}
%   F.\ Lastname1, F.\ Lastname2, and F.\ Lastname3,
%   ``Title of your INTERSPEECH 2022 publication,''
%   in \textit{Interspeech 2022 -- 23\textsuperscript{rd} Annual Conference of the International Speech Communication Association, September 18-22, Incheon, Korea, Proceedings, Proceedings}, 2022, pp.~100--104.
% \end{thebibliography}

\end{document}